\documentclass[twocolumn,showpacs,prd,amsfonts]{revtex4}

\usepackage{bm}

\begin{document}
\title{Are there hyperentropic objects ?}

\author{Jacob D. Bekenstein}\email{bekenste@vms.huji.ac.il}
\homepage{http://www.phys.huji.ac.il/~bekenste/}
\affiliation{Racah Institute of Physics, Hebrew University of
Jerusalem\\ Givat Ram, Jerusalem 91904 ISRAEL\\and\\
Jefferson Physical Laboratory, Harvard University\\ Cambridge, MA 02138}

\date{\today}
\begin{abstract} 
  By treating the Hawking radiation as a system in thermal equilibrium, Marolf and R. Sorkin have argued that hyperentropic objects (those violating the entropy bounds) would be emitted profusely with the radiation, thus opening a loophole in black hole based arguments for such entropy bounds.  We demonstrate, on kinetic grounds, that hyperentropic objects could only be formed extremely slowly, and so would be rare in the Hawking radiance, thus contributing negligibly to its entropy.  The arguments based on the generalized second law of thermodynamics then rule out weakly self-gravitating hyperentropic objects and a class of strongly self-gravitating ones.  
\end{abstract}


\maketitle 

\section{Introduction\label{intro}}

Hyperentropic objects (HEOs) were conjectured by Marolf and R. Sorkin (MS)~\cite{MS} as a challenge to the holographic bound~\cite{thooft,susskind} and the universal entropy bound~\cite{bek81,bek94,bek99}.   The holographic bound claims that within a closed 2-surface of area $A$ the entropy $S$ is restricted by
(henceforth we set $G=c=1$ but display $\hbar$) 
\begin{equation}
S\leq A/4\hbar.
\label{holographic}
\end{equation}
The universal entropy bound maintains that the entropy contained in an \emph{isolated} and \emph{weakly} self-gravitating object with total proper energy $E$ and radius $R$ is restricted by
\begin{equation}
S\leq 2\pi ER/\hbar.
\label{universal}
\end{equation}
For weak self-gravity systems Eq.~(\ref{universal}) implies Eq.~(\ref{holographic}).   Eqs.~(\ref{holographic}) and (\ref{universal}), and Bousso's covariant entropy bound~\cite{bousso,bousso-rev} from which these two follow---each under specific assumptions---have been widely regarded as constraining the form of the fundamental physical theory.     

Although no known systems subject to accepted physics exceed bounds~~(\ref{holographic}) and (\ref{universal}), it has been of interest to derive them from \emph{gedanken} experiments~\cite{bek81,bek94,susskind,bek00} in which the generalized second law of thermodynamics (GSL)~\cite{bek72} is challenged by hypothetical systems disobeying one of the said bounds.  An opposing school of thought~\cite{UW,wald} maintains that the GSL holds automatically, so that entropy bounds cannot be inferred from situations where the law seems to be violated.  Thus for example, the derivation of bound~(\ref{universal}) which considers the lowering of an entropy bearing object into a black hole (BH)~\cite{bek81}, and then looks at the impact on the GSL, was countered with the suggestion~\cite{UW} that quantum buoyancy near the BH would modify the energetics of the process sufficiently to obviate any conflict with the GSL without requiring the help of an entropy bound.  Later work found that the buoyancy is significant only when the object is a proper distance from the horizon of order of its own height~\cite{bek82,bek94}.  In this near region the thermal ambience which buoys it up is not well described by the fluid model of Ref.~\onlinecite{UW}.  A  wave-scattering based buoyancy calculation~\cite{bek99} which thus complemented older derivations~\cite{bek82,bek94} again recovered bound~(\ref{universal}) by invoking the GSL.  Regrettably the controversy regarding the relation between the GSL and entropy bounds did not then abate. 

At any rate, quantum buoyancy is irrelevant for an object allowed to \emph{fall freely} into a BH.   This made possible an alternative argument for a bound on a \emph{weakly} self-gravitating object's entropy which is just a factor of ${\cal O}(10)$ above (\ref{universal})~\cite{MG94,erice01,bek04}.  In the new argument the object is dropped from such a distance that the energy drawn from the hole by the Hawking radiance during the whole infall is balanced by the energy (at infinity) brought in by the object---the BH is left unchanged.  By the GSL the entropy carried by the object (the entropy lost to the hole) must be dominated by that of the Hawking radiance emitted concurrently.  This last depends on the number of (massless) species emitted, but for reasonable numbers the said weak version of bound (\ref{universal}) is recovered.

In their critique of this second derivation, MS contend that were the object's entropy $S$ to exceed the bound by however large a factor, a confrontation with the GSL would still be avoided because the outgoing Hawking radiation would then contain a multitude of similar objects, and their associated entropy would more than balance the entropy influx associated with the dropped object.  Now as a rule composite objects with energies above Planck's will not show up in the Hawking radiation~\cite{MG94}, being strongly suppressed by the usual Boltzmann factor.  The MS proposal gets around this hurdle by relying on the very large numbers of internal states appropriate for objects that violate bound~(\ref{universal}).  

MS's argument may be paraphrased as follows.  In a canonical ensemble of temperature $T$, the probability to find \emph{some} state of energy $E$ of an object is $p=Z^{-1}\,e^{-E/T}$, where $Z$ is the total partition function which can be written $e^{-F_t/T}$ with $F_t$ the total free energy.  The multiplicity of states of the object having energy $E$ can be expressed in terms of the object's internal entropy $S$ as $e^{S}$.  Hence the probability to find the object in any state is $p=e^{(F_t-F)/T}$ where $F=E-TS$ denotes its free energy at temperature $T$.  If the object in question is bosonic, there are
\begin{equation}
g(n)={(n+e^S-1)!\over n! (e^S-1)!}
\end{equation}
ways for $n$ identical objects to appear, with the corresponding probability being $p(n)=g(n)\, e^{(F_t-nE)/T}$.   When $F<0$, or $T>E/S$, $p(n)$ \emph{grows} with $n$ at first peaking for large $e^S$ at $n\sim e^{|F|/T}$.  A similar result is obtained for fermions.  Thus the ensemble may become dominated by objects that have $F<0$.   

MS apply this to objects in the thermal ambience of a Schwarzschild BH of mass $M$ and temperature 
\begin{equation}
T_{BH}=\hbar/(8\pi M).
\label{T}
\end{equation}
Regarding the typical object as spherical with radius $R$, and parameterizing its entropy by $S=\alpha R E/\hbar$ with $\alpha$ dimensionless, one
finds that $F=E(1-\alpha R/8\pi M)$. In the infall derivation of bound~(\ref{universal}) one is particularly interested in the case $R\, \hbox{\raise2pt\hbox{$<$} \kern-11pt\lower2pt\hbox{$\sim$}}\,
 2M$.  Accordingly, if $\alpha \gg 2\pi$, $F<0$. Therefore, MS contend that objects of this same kind may be emitted profusely with the Hawking radiation, and that the consequent entropy outflux will overpower the influx from the single infalling object.  
In particular, MS conclude that the existence of HEOs with however large an $\alpha$ cannot be ruled out by appealing to the GSL; they reach a similar conclusion in regard to objects that exceed the holographic bound.  The arguments detailed below, focusing mainly on the case $\alpha\gg 2\pi$, refute these conclusions 

Whatever the merit of MS's argument for true thermodynamic equilibrium, it is found wanting for the Hawking radiation from the viewpoint of \emph{kinetics}.   The success of simple statistical-thermodynamic formulae in describing the spectrum and statistics of Hawking emission into empty space should not obscure the fact that the emission is a nonequilibrium process, e.g. entropy increases during Hawking emission~\cite{bek75,page2}.  Thus the question at what \emph{rate} are HEOs formed in the Hawking radiation is paramount; \emph{equilibrium} formulae may not accurately reflect their true abundance.    

In this respect it is important to realize that HEOs must be composite objects.  For if we denote a HEO's energy and radius by $E_*$ and $R_*$, respectively, then as for any quantum system, $E_*R_*/\hbar>1$ (radius exceeds Compton wavelength).  Now if $\alpha\gg 2\pi$, we have $S_*\gg 2\pi$, meaning the HEO has a large number $e^{S_*}\gg 535$ of internal states.   But an elementary particle has just a few internal (spin) states, so the HEO is composite.  Of course in that case $E_*R_*/\hbar\gg 1$ because $E_*$ is many times the energy of a single component, and that component's Compton length must bound $R_*$ from below.   Thus $e^{S_*}$ is actually much larger than appears from the above argument.

Because an HEO is composite, it makes sense to ask what does it take for one to assemble spontaneously out of \emph{equilibrium} thermal radiation at temperature $T$ ?   Now a volume $V$ of such radiation has mean energy 
\begin{equation}
\langle E\rangle=V N(\pi^2 T^4/30 \hbar^3),
\label{mean}
\end{equation}
where $N$ is the effective number of helicity species (2 for the photon, $7/8$ for each chiral neutrino species, etc.).  In the real world $N={\cal O}(10)$; the arguments below work for $N< 180\pi^2 \approx 1776$.  Thus at the MS temperature 
\begin{equation}
T_0=E_*/S_*=\hbar/(\alpha R_*),
\label{MS}
\end{equation}
a mean energy $E_*$ of radiation is spread over a lengthscale $V^{1/3}=\alpha(30/\pi^2 N)^{1/3} S_*{}^{1/3}R_*\gg R_*$.  (By contrast, the energy of a typical massless quantum---one with wavelength $\lambda\sim \hbar/T$---is contained in a volume $\sim \lambda^3$.)  Thus according to the usual understanding, it will take a long time for fluctuations to concentrate energy $E_*$ within a scale $R_*$.  Note that one cannot rely on gravitation to concentrate the energy.   Even if the to-be-formed HEO is strongly self-gravitating, $V\gg (4\pi/3)R_*{}^3$ so the energy in $V$ is still weakly self-gravitating, and gravity is powerless to concentrate it. Thus at $T\, \hbox{\raise2pt\hbox{$>$} \kern-11pt\lower2pt\hbox{$\sim$}}\, T_0$ HEO self-assembly is very slow.

Another way to say this is that for $T\, \hbox{\raise2pt\hbox{$>$} \kern-11pt\lower2pt\hbox{$\sim$}}\, T_0$, the equilibrium probability $P(E_*)$ to find radiation energy $E_*$ in a volume $4\pi R_*{}^3/3$ is tiny.  One calculates this by noting that since each mode of radiation is statistically independent, the total energy in all the many modes must, by the central limit theorem, be distributed normally about the mean $\langle E\rangle$ given by Eq.~(\ref{mean}) with the corresponding variance whose textbook value is $(\Delta E)^2 = T^2\partial \langle E\rangle/\partial T$, or \begin{equation}
(\Delta E)^2=V N(2\pi^2 T^5/15 \hbar^3).
\label{variance}
\end{equation}
Putting $V=4\pi R_*{}^3/3$ and defining $\chi\equiv T/(\alpha T_0)$, we find
\begin{equation}
P(E_*)\propto \exp\Big[-{\pi^3\over 90}{[N\chi^4-(45/2\pi^3)(R_* E_*/\hbar)]^2\over N\chi^5}\Big].
\label{suppression}
\end{equation}
At $T\ \hbox{\raise2pt\hbox{$>$} \kern-11pt\lower2pt\hbox{$\sim$}}\ T_0$ we have  $\chi\ll 1/(2\pi)$ so that the $N\chi^4$ term in the square is negligible  (recall that $R_* E_*/\hbar\gg 1$).  Thus in this temperature range the probability is that for which $E_*=\langle E\rangle$ multiplied by an exponentially small factor $\ll\exp[-(180\pi^2/N)(R_* E_*/\hbar)^2]$.  In fact, this suppression is stronger than the enhancement factor $e^{S_*}=\exp(\alpha R_* E_*/\hbar)$ invoked by MS. Hence, although equilibrium thermal radiation at a temperature substantially above $T_0$ may eventually contain HEOs, it will take a long time for a fluctuation large enough to make the first HEO.  

This remains true until $T$ is raised so much that $N\chi^4$ approaches $0.73 (R_* E_*/\hbar)$; only then, as the argument of the exponential in (\ref{suppression}) become small, will exponential suppression cease.  This corresponds to $E_*\approx \langle E\rangle $ with $V=4\pi R_*{}^3/3$, and happens when $T\approx T_*$, where the new temperature scale,
\begin{equation}
T_*\equiv \alpha N^{-1/4} (R_* E_*/\hbar)^{1/4}\, T_0,
\label{Tstar}
\end{equation}
is well above $T_0$.    The value (\ref{Tstar}) for $T_*$ for strong self-gravity HEOs is changed by only a factor close to unity  because all stable self-gravitating objects---BHs aside--involve red-shift and length contraction factors of ${\cal O}(1)$.

Let us apply these results to HEO formation in the thermal ambience of a Schwarzschild BH. Suppose first that $R_*\ll 2M$.  Then $E_*\ll M$ even if the object is strongly self-gravitating, so it a small perturbation off the BH background.  Now at proper distance $\ell\ll 2M$ from the horizon, the \emph{local} temperature is $T= \hbar/2\pi \ell$~\cite{li,bek94,zas}; therefore, the highest available temperature, $T_h$, is the local temperature a proper distance $\ell \approx R_*$ from the horizon at which point the bottom of the HEO almost touches it.  Thus $T_h\approx \hbar/2\pi R_*$, or
\begin{equation}
T_h\approx (2\pi)^{-1} N^{1/4} (E_* R_*/\hbar)^{-1/4}\,T_*\ll T_*.
\label{first}
\end{equation}

Accordingly, since the available temperatures are well below $T_*$, the exponential suppression reemerges, and HEOs with $R_*\ll 2M$ will form extremely slowly in the Hawking radiance; since that streams rapidly out, they should occur only rarely.   Of course it could be claimed, in the spirit of Hawking's original derivation, that the HEOs in question already emerge whole from the BH in late time modes of some effective quantum field.  But such HEOs would be \emph{quasiparticles} of a composite field; as such they would be subject to decay.  The prospect for a treatment by nonequilibrium quantum field theory of HEO Hawking emission, which could clarify this issue, seems remote.  In any case, the long time nature of the Hawking modes does not favor the claim since it refers to global time, while the relevant time for kinetic processes is the more rapidly elapsing local time.  Our conservative conclusion is that $R_*\ll 2M$ HEO's contribute insignificant entropy outflow.

An HEO with $R_*\sim 2M$ \emph{cannot}, by virtue of its size, come from the BH itself, but it might be formed in the Hawking radiation at some distance from the horizon.   The largest temperature available for it is thus about $T_{BH}$, namely $T_h\sim  \hbar/4\pi R_*\ll T_*$, which is again too low for significant HEO presence.   We remark that, if strongly self-gravitating, the HEO in question would have an $E_*$ amounting to a significant fraction of $M$, and would thus constitute a sizable energy burden on the radiation, and ultimately on the hole.  This fact alone should further suppress the abundance of strongly self-gravitating HEOs with $R_*\sim 2M$.

The last case is $R_*\gg 2M$.  Here we can only contemplate weakly self-gravitating objects; otherwise $E_*$ would be comparable or exceed $M$, and such objects would not appear in the radiation.    What is $T_h$ here ? Although  $T_{BH}$ is the temperature that characterizes the spectrum and statistics of the Hawking radiation, the \emph{density} of radiation energy is characterized by a lower ``radiation temperature'' because at distance $r\gg 2M$ from the hole ($r$ is the usual Schwarzschild radial coordinate and approximates radial distance), the radiation energy density has been diluted by the factor $(2M/r)^2$.  Accordingly, $T_h=(2M/r)^{1/2}T_{BH}$.  Putting $r> R_*$  we have
\begin{equation}
T_h<\Big({\surd N\over 16\pi^2}{\hbar/E_*\over 2M}\Big)^{1/2}\, T_*.
\label{third} 
\end{equation}
Thus significant presence of $R_*\gg 2M$ HEOs in the radiation requires that the HEO \emph{Compton wavelengths} be fairly large compared to the hole's radius $2M$.

Motivated by the compositeness of HEOs, we now make the physically reasonable hypothesis that there is a largest HEO radius for given $E_*$, $R_{\rm max}(E_*)$, and that it does \emph{not} decrease with increasing $E_*$.  Of course, gravity can make the more massive of a family of composite objects smaller than the lighter ones, e.g. neutron stars, but as already mentioned, for $R_*\gg 2M$ we are restricted to weak self-gravity HEOs, so that this complication is excluded. 

Imagine dropping a HEO candidate with mass $E_1$ and radius $R_1=R_{\rm max}(E_1)$ into a Schwarzschild BH of radius $M=\zeta R_1$; here $\zeta$ is chosen a few times unity, so that the object can fall into the BH without being torn up.  Note that the candidate must have weak self-gravity; otherwise $E_1$ would not be small compared to $M$, so that the procedure would severely perturb the BH.  The drop is performed in harmony with the prescriptions of Ref.~\onlinecite{erice01}. 

Now by the results mentioned following our Eq.~(\ref{first}), HEOs smaller than $R_1$, comparable to it, or just a few times larger are not found in the said BH's radiation.  What about HEOs with $R_*> 2M>R_1$ and some allowed energy $E_*$ ?  We have $R_{\rm max}(E_*)\geq R_*>R_1 $.  By our hypothesis, the HEO energies must satisfy $E_*> E_1$.  Hence, $2M>R_1>\hbar/E_1>\hbar/E_*$.  Thus by result (\ref{third}), these bigger HEOs are also essentially absent from the Hawking radiation of our BH.  In summary, HEOs (other than the dropped one), whether weakly or strongly self-gravitating, play a \emph{negligible} role in the energy and entropy balances of the envisaged \emph{gedanken} experiment. 

By the dropping protocol~\cite{erice01}, the mass loss of the BH to Hawking emission of elementary quanta during the infall is designed to just compensate the mass gain from the dropped object---the BH does not change overall.  Were the emission reversible, the radiated entropy  would be just $E_1/T_{BH}$.  Irreversibility and the curvature of the spacetime make the entropy larger by a factor
 $\nu=1.35$--$1.64$~\cite{page2}.  Thus the overall change in
world entropy is     
\begin{equation}
\delta S_{\rm tot} =  \nu E_1/T_{BH} - S_1 . 
\label{netchange} 
\end{equation} 
Then by the GSL
\begin{equation}  
S_1 < 8\pi\nu\zeta R_1 E_1/\hbar,   
\label{newbound}   
\end{equation} 
i.e., $\alpha<8\pi\nu\zeta$~\cite{MG94,erice01,bek04}.   Hence the candidate with $R_1=R_{\rm max}(E_1)$ cannot have very large $\alpha$ values, and is not an HEO as defined here.  Let us discard it and other candidates with $R=R_{\rm max}(E)$, and repeat our argument using the narrowed HEO candidate strip in the $E_*-R_*$ plane.    The end result is that by the GSL, objects with weak self-gravity ($R_*\gg 2E_*$) cannot have $\alpha$'s  larger than a few times $2\pi$ [weak version of bound~(\ref{universal})].  

For objects with weak self-gravity, the traditional holographic bound~(\ref{holographic}) follows immediately from our last result because $A_1\gg 8\pi E_1 R_1/\hbar$ for weak self-gravity.  Or it can be proved directly by considering the infall protocol of Ref.~\onlinecite{bek00} and allowing for the possibility of Hawking radiation HEOs as done here.

Can the holographic bound be violated by a strongly self-gravitating object ?  Not by a large factor as the following rough argument shows.  Again denote the object's radius, energy and entropy by $R_1$, $E_1$ and $S_1$, respectively.  Strongly gravitating means $R_1=2\xi E_1$ with $\xi$ just a few times unity.  Let us repeat the drop into a BH as detailed in the last paragraphs.  To avoid strong perturbation of the BH we take $\zeta$ sufficiently large; since $M=2\zeta\xi E_1$, $\zeta={\cal O}(10)$ will suffice.  The argument against Hawking radiation HEOs goes as above; in particular, for $R_*\gg 2M$ we only have to consider weakly self-gravitating HEOs, and may again employ the argument based on $R_{\rm max}(E_*)$.   Inequality (\ref{newbound}) is recovered and translates into  $S_1 < (4\nu\zeta/\xi)\pi R_1{}^2/\hbar$.  Evidently, any violation of bound (\ref{holographic}) by weak self-gravity objects is by a factor smaller than $(4\nu\zeta/\xi)$, i.e., by no more than an order of magnitude.   

The original holographic bound (\ref{holographic}) can be recovered for a strongly self-gravitating object massive on Planck scale by revamping Susskind's original GSL argument for that bound~\cite{susskind}.   That argument envisages adding entropy-free matter to the candidate object to coax it into collapsing to a black hole.  MS dismiss the argument by suggesting that the collapse may lead, not a black hole, but to an ephemeral thermal fluctuation that soon disperses.  But there is, in fact, a \emph{gedanken} experiment in which a preexisting black hole of small dimensions and negligible entropy is used to catalyze the collapse of the candidate object to a black hole while the whole system is confined to a cavity~\cite{bek00}.  The slow formation of HEOs in the auxiliary hole's radiation allows the system to reach the partial thermal equilibrium state required for the deduction of the bound before HEOs become significant for the entropy balance.  Bound (\ref{holographic}) is then recovered.

We have established here under mild assumptions  that any overshoot of the universal and holographic bounds is by no more than a factor of ${\cal O}(10)$.  Accordingly, HEOs surpassing either of the bounds by arbitrarily large factors, a possibility suggested by MS, would be inconsistent with the GSL.  It should be stressed that the method described here is not particularly well suited to derive \emph{optimal} entropy bounds.  We believe bounds~(\ref{holographic}) and (\ref{universal}) are valid as stated, each within the specified limitations~\cite{bousso-rev}. 

I thank Larry Ford for valuable comments. This research was supported by the Israel Science Foundation (grant No. 694/04).

\end{document}